\documentclass[aps,pra,twocolumn,amsmath,amssymb,longbibliography,superscriptaddress]{revtex4-2}
\pdfpageattr {/Group << /S /Transparency /I true /CS /DeviceRGB>>}
\usepackage{graphicx}
\usepackage{color}
\usepackage[hidelinks]{hyperref}
\usepackage{orcidlink}
\usepackage{amsmath}
\usepackage{amsthm}
\usepackage{physics}
\usepackage{amsfonts}
\usepackage{times,txfonts}

\hypersetup{colorlinks=true, linkcolor=blue, citecolor=blue, urlcolor=blue}

\begin{document}
\author{George Mihailescu\,\orcidlink{0000-0002-0048-9622}}
\email[]{george.mihailescu@ucdconnect.ie}
\affiliation{School of Physics, University College Dublin, Belfield, Dublin 4, Ireland}
\affiliation{Centre for Quantum Engineering, Science, and Technology, University College Dublin, Dublin 4, Ireland}

\author{Steve Campbell\,\orcidlink{0000-0002-3427-9113}}
\email[]{steve.campbell@ucd.ie}
\affiliation{School of Physics, University College Dublin, Belfield, Dublin 4, Ireland}
\affiliation{Centre for Quantum Engineering, Science, and Technology, University College Dublin, Dublin 4, Ireland}
\affiliation{Dahlem Center for Complex Quantum Systems, Freie Universit\"{a}t Berlin, 14195 Berlin, Germany}

\author{Karol Gietka\,\orcidlink{0000-0001-7700-3208}}
\email[]{karol.gietka@uibk.ac.at}

\affiliation{Institut f\"ur Theoretische Physik, Universit\"at Innsbruck, Technikerstra{\ss}e\,21a, A-6020 Innsbruck, Austria} 

\title{Uncertain Quantum Critical Metrology: From Single to Multi Parameter Sensing}


\begin{abstract}
Critical quantum metrology relies on the extreme sensitivity of a system's eigenstates near the critical point of a quantum phase transition to Hamiltonian perturbations. This means that these eigenstates are extremely sensitive to \emph{all} the parameters of the Hamiltonian. In practical settings, there always exists a degree of experimental uncertainty in the control parameters---which are approximately known quantities. In realistic settings, there is always some degree of uncertainty in the control parameters used to tune the system to criticality. These uncertainties, while not the target of estimation, can significantly affect the attainable precision—effectively acting as nuisance parameters in the estimation process. Despite being a practically relevant source of noise, their impact on critical metrology has been largely overlooked. In this work, we present a general framework that interpolates between single- and multi-parameter estimation settings, enabling a systematic analysis of how such uncertainties influence sensitivity. We apply this framework to the paradigmatic transverse field Ising and Lipkin-Meshkov-Glick models, explicitly demonstrating how uncertainty in control parameters affects the metrological performance of critical sensors. For finite-size systems, we identify a fundamental trade-off between robustness to uncertainty and the ability to retain a quantum advantage at the critical point. Our results contribute to a deeper understanding of the practical limitations of critical quantum metrology and provide a route toward its more resilient implementation.

\end{abstract}
\date{\today}
\maketitle



\section{Introduction}Critical quantum metrology~\cite{zanardi2007criticalscaling,zanardi2008informationgeometry,zanardi2008criticalityresource} aims to leverage the increased sensitivity to parameter variations for a system close to a quantum phase transition (QPT) in order to achieve high precision measurements~\cite{zakrz2018criticalb,garbe2020criticalmetrology,2020saulo,witko2021CQMferro,abol2021criticalityglobalsensing,e23101353,Gietka2022understanding,Pavlov_2023,plenio2022PRX,Garbe_2022,gietka2023overcoming,Marti,paris2023jointloos,paraoanu2023parametriccritical,stefan,salvia2023CQMrealtimefeedback,Gietka2024combining,gietka2024tempcqm,mihailescu2024quantumsensingnanoelectronicsfisher,li2024quantummetrologicalcapabilityprobe,alushi2024optimality,relation2024zhang,zicari2024criticalityamplifiedquantumprobingspontaneous,alushi2024collectivequantumenhancementcritical,paris2024spinchain}. This can be quantified by directly translating the sensitivity of Hamiltonian parameters to be estimated through the Cramér-Rao bound, which relates the quantum Fisher information (QFI) with the estimation uncertainty~\cite{caves1994statisticaldistance}. When the parameter of interest can be related to the relevant order parameter of the QPT, such increased sensitivity can be achieved. However, an important caveat is that all other parameters of the problem must be precisely known for the framework of single parameter estimation theory to be applicable. In general, this assumption is rather stringent. Typically, some degree of experimental uncertainty or noise is present~\cite{noisy2011metrology,preskill,Super_Heisen_Experimental}. In the context of quantum critical systems, while high precision experiments are possible~\cite{rydberg_metrology}, as complexity and system size increase—a requisite for critical quantum many-body systems—the resources required to perfectly control and characterize these systems, such as time, also increase~\cite{zakrz2018criticalb,garbe2020criticalmetrology,Gietka2021adiabaticcritical}. Furthermore, universality is a hallmark of systems approaching their critical point~\cite{heyl2015universality,plenio2015universalityQRM}, making the system dependent on many, if not all, Hamiltonian parameters precisely where the expected critical enhancement can be achieved. It is therefore intuitive that any uncertainty in at least one of these parameters is expected to negatively affect the QFI and thus deteriorate any potential quantum advantage. While the impact of noise on standard metrological techniques has been extensively studied~\cite{frequencystand1997decoh,noisey2011durkin,wscher2011Natphysnoise,demko2012elusigeHL,variatioanl2012noisey,mukhopadhyay2024saturableglobalquantumsensing,chaves2013noisey,maccone2014noiseagainst}, its effect on the performance of critical sensing has received considerably less attention~\cite{garbe2020criticalmetrology,gietka2024tempcqm,alushi2024optimality}. Specifically, it has been shown that thermal noise can enhance the performance of critical sensors~\cite{gietka2024tempcqm}, much like the noise arising from violations of adiabatic conditions~\cite{dynamical2021Cai,Gietka2022understanding,Garbe_2022}. 

Despite recent advances, a crucial practical challenge in critical quantum metrology remains largely underexplored: the detrimental effect of uncertainty in control parameters on the achievable sensitivity near critical points. These control parameters---used to tune the system toward criticality---are not the target of estimation, yet their fluctuations significantly influence the probe's dynamics and measurement outcomes. In this context, they naturally assume the role of nuisance parameters~\cite{Suzuki_2020_nuisance}. Namely, variables that are not of direct interest but nonetheless impact the precision of the parameter being estimated. This is especially problematic in critical metrology, where proximity to a critical point is essential for exploiting the system’s diverging susceptibility. Unlike standard quantum metrology protocols, which often do not require precise control over external parameters, critical metrology is uniquely vulnerable to such uncertainties. Thus, understanding and mitigating the effects of nuisance parameters arising from imperfect control is essential for realizing the full potential of criticality-enhanced quantum sensing.


A more appropriate treatment of critical metrology requires a methodology that relaxes the core constraint of the single parameter framework, i.e., that all parameters are assumed to be perfectly known~\cite{fresco2022multiparameterQCM,mihailescu2023multiparameter,PhysRevA.106.012424,mondal2024multicriticalquantumsensorsdriven}. In addition to providing a more realistic description of the actual achievable precision, this approach also allows for mitigating the necessity of complex and resource-intensive experimental control, ultimately enabling the identification of regions where a true quantum advantage is achievable. A candidate framework for such a purpose is multiparameter quantum metrology~\cite{ragy2016compatibility,Liu_2020,fresco2022multiparameterQCM,mihailescu2023multiparameter,2023kolodyn,zoller2023multiparameter}, where the aim is typically to estimate multiple parameters, in principle simultaneously. Clearly, within this framework, we can still assume there is a central parameter we wish to infer, and additionally, there are other relevant parameters about which we have no {\it a priori} information. Contextually, in terms of critical metrology, the QPT is driven by a single, experimentally controllable system parameter \(\beta\), and our objective is to infer the desired system parameter \(\alpha\). The multiparameter paradigm represents the situation in which the driving (control) parameter \(\beta\) is unknown. However, typically the driving parameter is an experimentally controllable knob and therefore is known within some finite window of resolution. It should be clear then that the multiparameter paradigm provides us with a worst-case scenario for the estimation of a central desired quantity, \(\alpha\), whereby we have a complete lack of knowledge regarding the value of the driving parameter, \(\beta\).

In this work, we present a general approach to quantum parameter estimation, interpolating between the two extreme scenarios of single and multi-parameter estimation, allowing for the proper bookkeeping of relevant errors~\cite{abol2021criticalityglobalsensing,paris2014cqmLMG,PhysRevA.106.012424}.We apply this formalism to quantum critical metrology and investigate how uncertainties in the driving parameter influence the ability to infer an unknown parameter and whether a relevant advantage is maintained. We consider the Ising and Lipkin-Meshkov-Glick models as two paradigmatic critical systems, showing that the multiparameter estimation framework predicts a catastrophic failure in the sensing capabilities of the probe in these models. By allowing for a degree of statistical uncertainty in the driving parameter, our framework demonstrates that a quantum advantage can still be achieved with critical probes even in the presence of such uncertainties.


\section{Parameter estimation in the presence of uncertainties}In the most general quantum sensing scenario, there are a set of unknown parameters \(\vec{x}\) to be estimated through suitable measurements of a quantum probe~\cite{QM2006lloyd,AQM2011LLoyd,liu2019quantum}. Information regarding this set of parameters is encoded in the state of the probe, given by the density matrix \(\hat{\varrho}\left(\vec{x}\right) = \sum_i \lambda_i\ket{\lambda_i}\bra{\lambda_i}\), with \(\lambda_i\) being the probability of the probe to occupy state \(|\lambda_i\rangle\) which depends on \(\vec{x}\). This information is determined by performing a large number of measurements. The precision of parameter estimation for a single measurement round is given by the quantum Cramér-Rao bound~\cite{caves1994statisticaldistance} (hat notation indicates a matrix):
\begin{align}
    \label{eq:MPCRB}
    \mathrm{Cov}\left[\vec{x}\right] \geq {\hat{\mathcal{I}}}^{-1},
\end{align}
{{which is a matrix inequality}}, lower bounding the precision of parameter estimation through the covariance matrix. The elements are \(\text{Cov}(\alpha,\beta) = \langle (\alpha - \langle \alpha\rangle )(\beta - \langle \beta\rangle )\rangle\), where \(\alpha\) and \(\beta\) label parameters from \(\vec x\). Entries of the quantum Fisher information matrix (QFIM) are given by:
\begin{align}
    \label{eq:QFIM_Entry}
    {\mathcal{I}}_{\alpha\beta} = \sum_{i,j}\frac{2\Re\left[\langle\lambda_i|\partial_\alpha \hat{\varrho}|\lambda_j\rangle\langle \lambda_j|\partial_\beta  \hat{\varrho} |\lambda_i\rangle\right]}{\lambda_i + \lambda_j},
\end{align}
where diagonal elements relate to the single parameter precision and are obtained by suitably maximizing over all possible positive operator-valued measures. The off-diagonal elements relate to the correlation between parameters. For pure states, the QFIM may be simplified to:
\begin{align}
    \label{eq:Pure_State_QFI}
    \mathcal{I}_{\alpha\beta} = 4\Re\left[\langle \partial_\alpha \psi|\partial_\beta \psi\rangle - \langle \partial_\alpha \psi|\psi\rangle \langle \psi| \partial_\beta \psi \rangle \right].
\end{align}
It is informative to consider the simple case where we have two parameters to be estimated, although the results readily extend to an arbitrary number of variables. From Eq.~\eqref{eq:MPCRB}, the attainable precision with respect to parameter \(\alpha\) is given by:
\begin{align}
\label{eq:alphavar}
    \big(\Delta \alpha\big)^2 \geq \frac{{\mathcal{I}}_{\beta\beta}}{{\mathcal{I}}_{\alpha\alpha}{\mathcal{I}}_{\beta\beta} - {\mathcal{I}}_{\alpha\beta}{\mathcal{I}}_{\beta\alpha}}.
\end{align}
where we note that \(\mathcal{I}_{\beta\alpha} = \mathcal{I}_{\alpha\beta}\). To contrast this, in the single parameter setting where \(\beta\) is assumed to be known precisely, Eq.~\eqref{eq:alphavar} reduces to \(\big(\Delta \alpha\big)^2 \geq 1/{\mathcal{I}}_{\alpha\alpha}\). {As the QFIM is a positive semi-definite matrix, it follows that precision in multiparameter estimation Eq.~\eqref{eq:alphavar} generally represents a reduction in sensitivity, compared with the precision attainable when all the other parameters are known~\cite{mihailescu2023multiparameter}, which gives simply  $\big(\Delta \alpha\big)^2 \ge 1/\mathcal{I}_{\alpha,\alpha}$.  The degradation in precision is controlled by correlations between multiple unknown parameters.} Thus, while the single-parameter estimation framework gives an optimistic lower bound on the precision, the multi-parameter case gives a pessimistic one. This becomes particularly evident when considering cases where the parameters of interest are related to one another. From Eq.~\eqref{eq:alphavar}, it is clear that in order for the multiparameter problem to be well posed, the QFIM must be an invertible quantity~\cite{2001parameterestimationsingular, taming_singular}, requiring that:
\begin{align}
    \text{det} \,{\hat{\mathcal{I}}}  = {\mathcal{I}}_{\alpha\alpha} {\mathcal{I}}_{\beta\beta} - {\mathcal{I}}_{\alpha\beta}^2 \neq 0.
\end{align}
Thus, when the QFIM is singular, i.e., \(\det \, {\hat{\mathcal{I}}} = 0\), the quantum Cramér-Rao bound ceases to be informative, effectively implying that the lowest achievable uncertainty for at least one parameter diverges. This typically signals that the parameters are not independently identifiable from the quantum state. For example, when the eigenstates depend only on a ratio of parameters, such as \(\alpha/\beta\), then changes in \(\alpha\) and \(\beta\) that preserve this ratio leave the state unchanged, making it impossible to estimate either parameter individually~\cite{taming_singular,mihailescu2025metrologicalsymmetriessingularquantum}. In critical metrology, this issue becomes especially pronounced, as the ground state near a critical point is highly sensitive to all system parameters. Consequently, incomplete knowledge of control parameters can render the system's metrological advantage ineffective, as the quantum state no longer provides distinguishable information about the parameter of interest.


In most relevant settings, control parameters are implemented with a given tolerance, making it more meaningful to consider bounding the attainable precision on \(\alpha\) when \(\beta\) is known within some finite interval of resolution. In full generality, we may model the sensitivity to parameter \(\alpha\) given some uncertainty in the driving parameter \(\beta\) by introducing a probability distribution, \(p(\beta)\), obtained from sampling the generated states for a given value of \(\beta\). Formally, this can be achieved by considering the probe state as an infinite mixture of states, each corresponding to a different value of \(\beta\). The corresponding sensitivity is then given by evaluating the effective single parameter QFI for this family of states, thereby explicitly accounting for imprecision or fluctuations in \(\beta\).

Critical metrology achieves increased sensitivity in parameter estimation by exploiting properties of eigenstates near the critical point of QPTs, and the optimal sensitivity is quantified using Eq.~\eqref{eq:Pure_State_QFI}. However, as we are dealing with a mixture of states---specifically a mixture of ground states---we require the ground state density matrix for a given choice of \(\beta\), defined as \(\hat{\varrho}\left(\alpha,\beta\right) = \ket{\psi_{GS}}\bra{\psi_{GS}}\). To sample the corresponding quantum state for a given choice of \(\beta\), we then weigh the ground state density matrix according to the probability distribution \(p(\beta)\). A natural choice is a Gaussian distribution, such that \(p\left(\beta\right) = \tfrac{1}{\sigma \sqrt{2 \pi}} \exp{{-\tfrac{1}{2}\left( \frac{\beta - \bar{\beta}}{\sigma} \right)^2}}\), yielding on average a final state of the form:
\begin{align}
    \label{Eq:Average_DM}
    \hat{{\varrho}}\left(\alpha,\bar \beta,\sigma\right) =  \int_{-\infty}^{\infty} d \beta \frac{\exp{-\frac{1}{2}\left(\frac{\beta - \bar{\beta}}{\sigma}\right)^2}}{\sigma\sqrt{2 \pi }} \hat{\varrho}\left(\alpha,\beta\right),
\end{align}
with $\bar{\beta}$ the average value of parameter \(\beta\), and \(\sigma\) the variance quantifying the (im)precision. Although Gaussian noise is not the most general form of noise---which can be system and substrate dependent---modeling noise using a Gaussian distribution is a standard theoretical practice in many cases and often captures the essential effects in quantum sensing. The corresponding sensitivity is then obtained from the QFI of the averaged state from Eq.~\eqref{Eq:Average_DM} with the use of Eq.~\eqref{eq:QFIM_Entry}, which we denote as ${\bar{\mathcal{I}}}_{\alpha\alpha}$. This scenario represents an interpolation between the various relevant settings. For \(\sigma \to 0\), we recover the single parameter estimation case where \(\beta\) is assumed to be precisely known; while \(\sigma \to \infty\) corresponds to the multi-parameter setting where \(\beta\) is unknown and, depending on the specific setting at hand, the QFIM may become singular. 


\begin{figure*}[!htb]
\centering
    \includegraphics[width=0.46\textwidth]{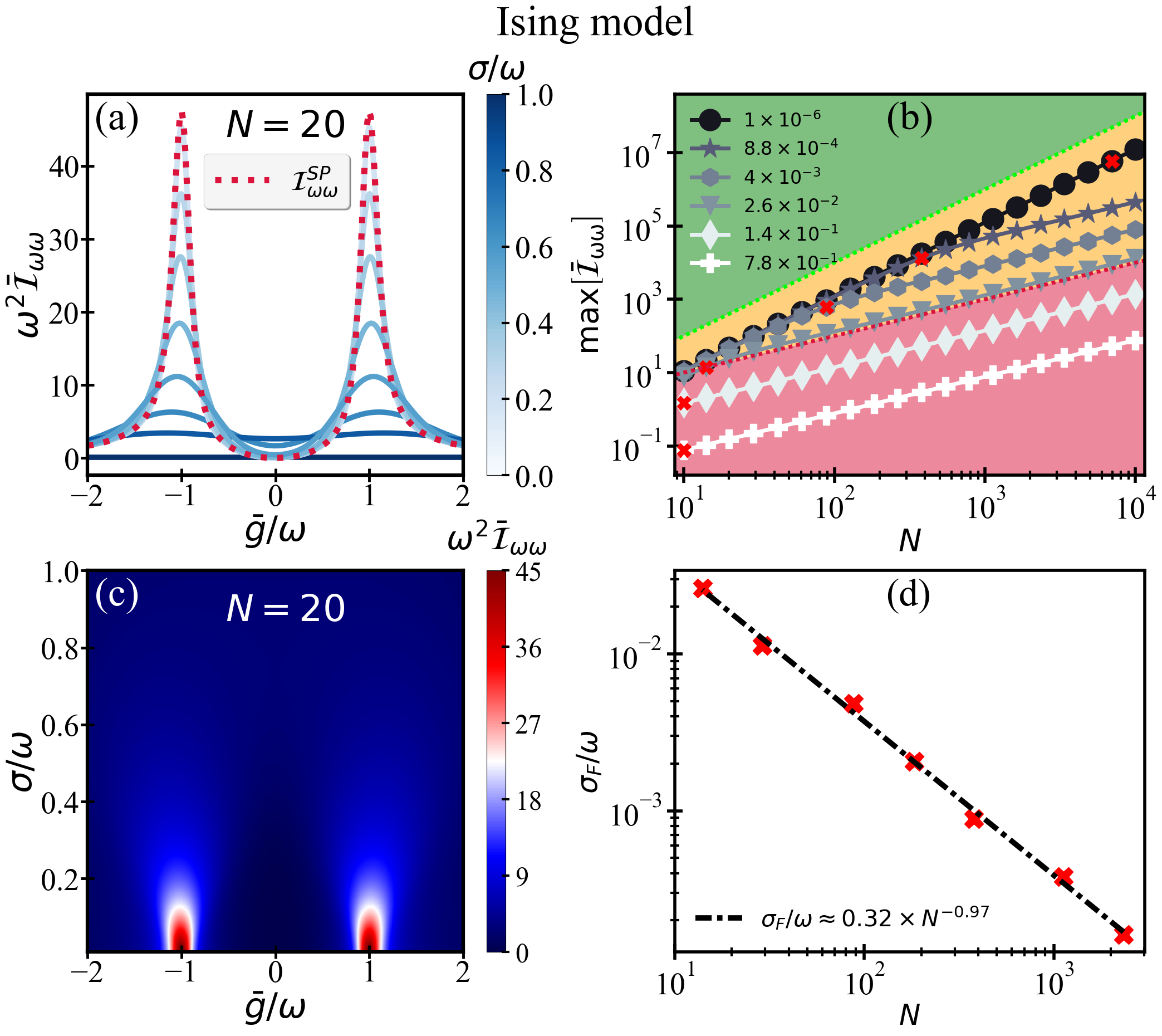}
    \hfill
    \includegraphics[width=0.46\textwidth]{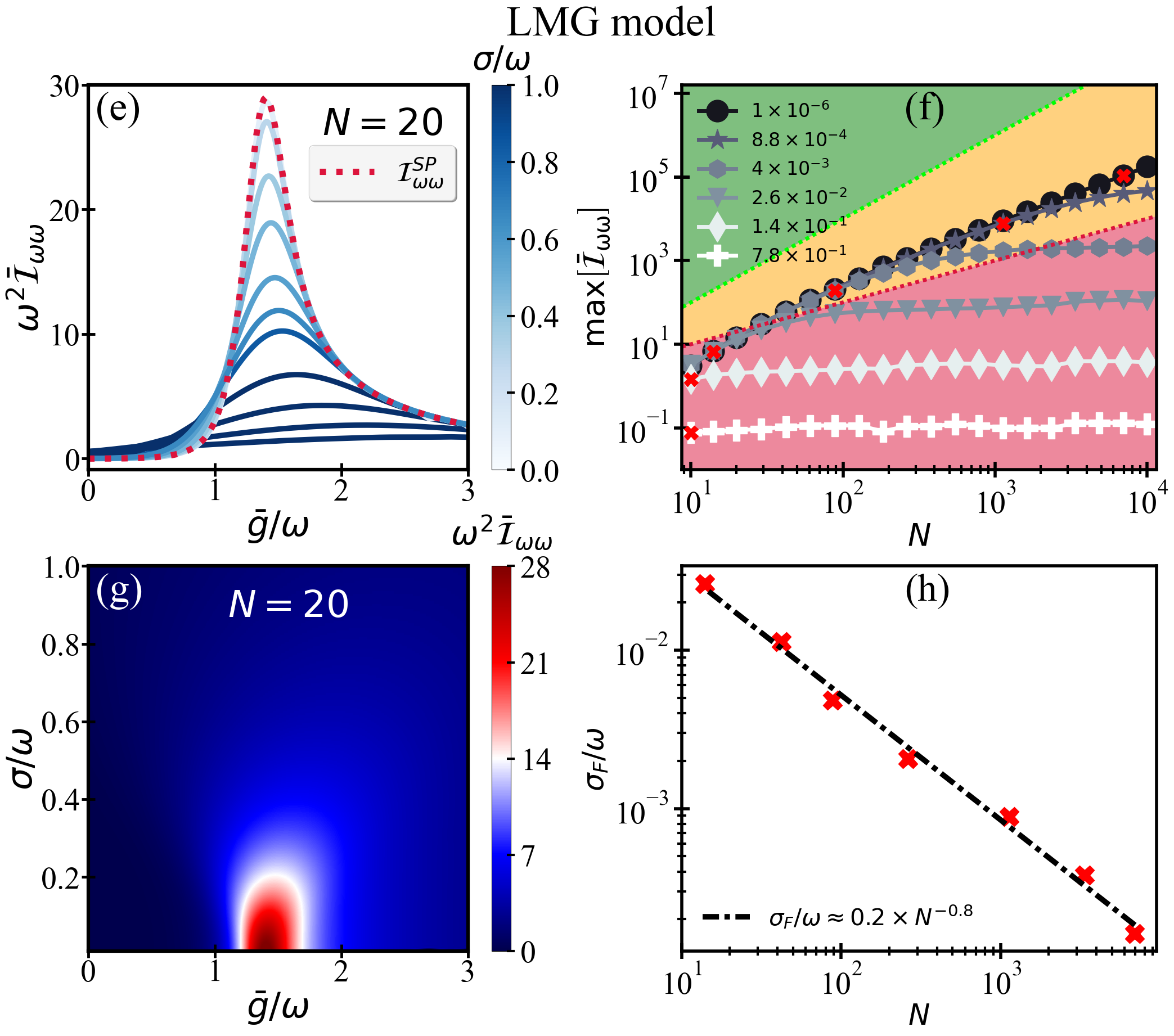}
	\caption{Left panel: \textit{Ising model}. Right panel: \textit{LMG model}. Sensitivity to the bare frequency $\omega$ with uncertainty in the coupling $g$ quantified by $\sigma/\omega$. In (a) and (c) as well as (e) and (g) the attainable sensitivity to the frequency $\omega$ is shown for various degrees of uncertainty for $N = 20$ spins. In (a) and (e), the dotted red line shows the ideal sensitivity $\mathcal{I}_{\omega\omega}^{SP}$ of the single parameter case when $g$ is perfectly known. Solid lines show attainable sensitivity for a given precision in $g$. (c) and (g) depict the full phase diagram for parameter sensitivity. (b) and (f) show scaling of the peak QFI with a given uncertainty, ${\bar{\mathcal{I}}}_{\omega\omega}$, in the coupling $g$ as a function of system size $N$. (d) and (h) depict the maximal uncertainty at which the QFI \emph{feels} the uncertainty of the control parameter $g$ as a function of system size $N$.}
	\label{fig:Ising_Results}
\end{figure*}

\section{Applications}We apply our framework to the general class of Hamiltonians given by:
\begin{align}
    \label{Eq:General_Ham}
    \hat{H} =  \omega \sum_{i}^{N} \hat{\sigma}_i^z - \sum_{i \neq j = 1}^{N} g_{ij} \hat{\sigma}_i^x \hat{\sigma}_j^x 
\end{align}
with a transverse field $\omega$ that we wish to estimate and $g_{i j}$ the interaction strength between the $i^\text{th}$ and $j^\text{th}$ spin, which we assume is controllable. We will consider two paradigmatic examples of the above: the transverse field Ising model (TFIM) when $g_{i j}\!=\!g$ is non-zero only for the nearest neighbors~\cite{PFEUTY197079TFIM,Stinchcombe_1973TFIM}, and the Lipkin-Meshkov-Glick (LMG) model when $g_{i j}\!\!=\!\!g/N$~\cite{LIPKIN1965188,oberthaler2010lmg,Li2022LMG,muniz2020LMG,vuletic2023LMG} (see Ref.~\cite{[{See Supplemental Material at }][{ for details.}]sup1} for toy-model Landau-Zener Hamiltonian). In both cases, it should be immediately evident that one can perform a simple rescaling of the Hamiltonian, and therefore the ground state will be dependent on the parameters' ratio, immediately implying that the QFIM is singular~\cite{mihailescu2023multiparameter}.

Explicitly, for the Ising limit assuming periodic boundary conditions, by utilizing a Jordan-Wigner mapping and transforming to momentum space, we can obtain an exact expression for the ground state of the many-body system, $\ket{\psi_{GS}}_N = \bigotimes_{k>0} \ket{\psi_{GS}}_k$, where $\ket{\psi_{GS}}_k = \cos{\theta_k/2}\ket{0}_k + \sin{\theta_k/2}\ket{1}_k$~\cite{[{See Supplemental Material at }][{ for details.}]sup1}. Exploiting this and leveraging the fact that the QFI is additive under a tensor product, such that ${\hat{\mathcal{I}}}_{\omega g}\left[ \ket{\psi_{GS}}_N \right] = {\hat{\mathcal{I}}}_{\omega g}\left[ \ket{\psi_{GS}}_{k1}\right] + {\hat{\mathcal{I}}}_{\omega g}\left[ \ket{\psi_{GS}}_{k2}\right] + \ldots + {\hat{\mathcal{I}}}_{\omega g}\left[ \ket{\psi_{GS}}_{kN}\right]$, we obtain a closed-form expression for the QFIM
\begin{align}
    \label{eq:TFIM_Exact}
  {\hat{\mathcal{I}}}_{\text{Ising}} = \sum_{k} \begin{pmatrix}
    \frac{g^2 \sin^2{k}}{\left(g^2 + \omega^2 - 2 g \omega \cos{k} \right)^2} & -\frac{g \omega \sin^2{k}}{\left(g^2 + \omega^2 - 2 g \omega \cos{k} \right)^2} \\ -\frac{g \omega \sin^2{k}}{\left(g^2 + \omega^2 - 2 g \omega \cos{k} \right)^2} & \frac{\omega^2 \sin^2{k}}{\left(g^2 +\omega^2 - 2 g \omega \cos{k} \right)^2}
\end{pmatrix},
\end{align}
with $k =  \pi(2n+1)/N$ and $n = 0,1,2,\ldots,N/2 - 1$, which is clearly singular for all $N$, irrespective of the distance to the critical point. Consequently, the estimation of the bare frequency $\omega$ is strictly prohibited when the coupling $g$ is unknown. The single-parameter sensitivity to the bare frequency $\omega$ corresponds to the first element of Eq.~\eqref{eq:TFIM_Exact}.

Allowing for uncertainty in the coupling, we construct a mixture of ground states sampled for a given value of $g$ according to Eq.~\eqref{Eq:Average_DM}. In Fig.~\ref{fig:Ising_Results}, we show the QFI for this averaged state, denoted ${\bar{\mathcal{I}}}_{\omega\omega}$, for the TFIM. Moreover, in Fig.~\ref{fig:Ising_Results}(a), we explicitly compare the ideal single-parameter sensitivity, ${\mathcal{I}}_{\omega\omega}$ shown in the dotted red line, with attainable sensitivity for a given uncertainty, in the range $\sigma/\omega \!=\! [0,1]$ from solid light blue to solid dark blue for $N = 20$ spins. Crucially, we note that for small uncertainty in $g$, the QFI retains the peak of the single-parameter case, and this peak sensitivity decreases as our uncertainty in $g$ increases. Notably, in the limit $\sigma \!\rightarrow\! \infty$, uncertainty in the coupling completely deteriorates sensitivity to the bare frequency $\omega$, recovering the expected multi-parameter result when the QFIM is singular. Fig.~\ref{fig:Ising_Results}(c) shows the full phase diagram of the QFI for $N = 20$ spins as a function of the variance, $\sigma/\omega$, where we clearly see that introducing uncertainty in our knowledge of the coupling strength reduces and smooths out the peak sensitivity. 

In the best-case scenario when the coupling $g$ is perfectly known, the peak sensitivity at the critical point $g = \omega$ as a function of system size scales as $\left( N^2 + N \right)/8\omega^2$. Herein lies the power of critical quantum metrology, whereby treating the system size $N$ as a resource for sensing, the QFI scales as $N^\gamma$ with an exponent which is typically $\gamma > 1$ with respect to parameter(s) which drive the phase transition. Such scaling represents an advantage over the independent (non-interacting) case where $\gamma = 1$. An important consideration is therefore if such an advantage is indeed attainable in the presence of relevant experimental imprecision. In Fig.~\ref{fig:Ising_Results}(b), we plot the scaling of the QFI in the presence of uncertainty, ${\bar{\mathcal{I}}}_{\omega\omega}$ as a function of system size $N$ for various degrees of imprecision in the control parameter $g$. The red region indicates scaling that is worse than linear ($\gamma < 1$), the yellow region indicates scaling better than linear but worse than quadratic ($2 > \gamma > 1$), and the green region indicates better than quadratic scaling ($\gamma > 2$). 
The red crosses indicate the point at which the QFI is affected by the imprecision: specifically, when the QFI no longer scales as in the ideal scenario when $g$ is known with absolute certainty, $\left(N^2 + N\right)/8\omega^2$. Importantly, we note from the red crosses that larger system sizes are less robust to relevant uncertainties—indicating an important experimental trade-off between attainable sensitivity endowed by larger system sizes and their robustness to relevant perturbations. This provides a crucial practical consideration in the design of criticality-enhanced sensors, whereby there is an optimal trade-off between the sensitivity offered by larger systems and their robustness to uncertainties. Furthermore, treating the definition of criticality to hold strictly for an infinite system, we note that in the asymptotic limit of system size any finite perturbation or uncertainty would necessarily prohibit scaling larger than $\gamma = 1$, leading to a constant factor improvement, consistent with Ref.~\cite{demko2012elusigeHL}. Fig.~\ref{fig:Ising_Results}(d) depicts the maximal uncertainty at which the QFI \emph{feels} the uncertainty of the control parameter $g$---the red crosses from (b), denoted $\sigma_F/\omega$, as a function of system size $N$ [for clarity we do not show all the generated data in (b), hence there are more red crosses in (d) than in (b)~\cite{[{See Supplemental Material at }][{ for details.}]sup1}]. From here we can see that smaller system sizes are more robust to the uncertainty than bigger systems. According to a numerical fit, $\sigma_F/\omega \approx 0.32 \times N^{-0.97}$ in the TFIM, indicating how precisely the control parameter $g$ has to be known as a function of $N$ to retain the same sensitivity as in the single-parameter paradigm. 

A qualitatively similar behavior can be demonstrated for the LMG model. The isometric all-to-all coupling means it is convenient to recast Eq.~\eqref{Eq:General_Ham} in terms of collective spin operators, $\hat S_i \equiv \sum_n \hat \sigma_i^n/2$ with $i \in \lbrace x, y, z \rbrace$~\cite{paris2014cqmLMG},
\begin{align}
    \hat H  =  \omega \hat S_z - \frac{g}{N}\hat S_x^2.
\end{align}
The LMG Hamiltonian experiences a second order QPT in the thermodynamic limit at $g = g_c = \omega$~\cite{vidal2007LMGtl}. This can be explicitly seen by using the Holstein-Primakoff transformation and applying the $N\rightarrow\infty$ approximation,
\begin{align}
    \hat H = \omega \hat a^\dagger \hat a - \frac{g}{4}\left(\hat a^\dagger + \hat a\right)^2.
\end{align}
In the thermodynamic limit, the eigenstates are squeezed Fock states of the non-interacting system $|n_\xi \rangle = \hat S(\xi)|n\rangle$, where $\hat S(\xi) = \exp{\frac{1}{2}\left(\xi^*\hat a^2-\xi\hat a^{\dagger2}\right)}$ is the squeezing operator and $\xi = \frac{1}{4} \ln\{1-g/\omega\}$ is the squeezing parameter. At the critical point, the eigenstates become infinitely squeezed, which is responsible for the non-analytic behavior at the critical point~\cite{gietka2022speedup}, while for finite-sized systems, this non-analytic behavior is smoothed out as a function of $N$~\cite{lmg2004vidal}.

Since in the thermodynamic limit the ground state is completely defined by the squeezing parameter $\xi$, the QFIM can be simply expressed as,
\begin{align}
\begin{split}
    \hat{\mathcal{I}}_{\text{LMG}} &= 2
    \begin{pmatrix}
[\partial_\omega \xi]^2 & [\partial_\omega \xi] [\partial_g \xi] \\
[\partial_\omega \xi] [\partial_g \xi] & [\partial_g \xi]^2 
\end{pmatrix} \\ &=
    \frac{1}{32}
    \begin{pmatrix}
\frac{\omega^2}{(\omega-g)^2\omega^2} & \frac{g \omega}{(\omega-g)^2\omega^2} \\
\frac{\omega g}{(\omega-g)^2\omega^2} & \frac{g^2}{(\omega-g)^2\omega^2} 
\end{pmatrix},
\end{split}
\end{align}
which, again due to the fact that the ground state depends on the ratio of parameters, is evidently singular regardless of the distance to the critical point. In order to calculate the QFI matrix elements for a finite-size system including uncertainties in the control parameter $g$, we resort to using numerical calculations. The uncertainty of the control parameter $g$ is incorporated in the same way as for the TFIM. The exemplary QFI with respect to $\omega$ for $N=20$ spins is presented in Fig.~\ref{fig:Ising_Results}(e) and the corresponding full phase diagram in the presence of uncertainty in Fig.~\ref{fig:Ising_Results}(g). We also plot the maximal attainable QFI as a function of the system size for various levels of uncertainty in the control parameter [see Fig.~\ref{fig:Ising_Results}(f)]. Interestingly, although the LMG model is an all-to-all interacting system, the QFI scales less than quadratically as in the case of nearest-neighbor interacting TFIM, $\gamma = 4/3$. Similar to the TFIM, we also notice that for every $N$ there exists a corresponding $\sigma_F / \omega$ at which the QFI stops following the ideal case single-parameter scaling, denoted by the red crosses. The numerical fit reveals that $\sigma_F/\omega \approx 0.2 \times N^{-0.8}$, shown in Fig.~\ref{fig:Ising_Results}(h).


\section{Conclusions}
In this work, we have presented a framework to characterize the effectiveness of critical quantum probes, explicitly accounting for uncertainties in control parameters, which we identify as {nuisance parameters} in the language of quantum estimation theory. Such uncertainties represent a serious obstacle to achieving quantum enhancement in critical sensing, as they influence the system dynamics without being directly estimated. Our framework systematically interpolates between the best- and worst-case scenarios, as captured by single- and multi-parameter estimation settings, thereby addressing the problem of the trade-off between the estimation errors of parameters of interest and nuisance parameters, particularly in the finite-sample regime.

While the primary power of critical quantum probes typically lies in the scaling of the quantum Fisher information with system size \( N \), we demonstrate that relevant uncertainties introduce a fundamental trade-off between sensitivity and robustness. Notably, in the asymptotic limit, any finite uncertainty in control parameters completely negates the scaling advantage of the quantum Fisher information---a direct manifestation of the influence of nuisance parameters on optimal estimation strategies. Our results therefore address the question of how much information about nuisance parameters is required to retain quantum advantage, providing a practical and theoretically grounded answer~\cite{Suzuki_2020_nuisance}.

We applied our framework to two paradigmatic models---namely the Lipkin-Meshkov-Glick model and the Transverse Field Ising Model---and developed an analytical approach to compute the full quantum Fisher information matrix and its determinant for any system that can be mapped to free fermions. This enables generalization to a wide range of many-body quantum systems, including \( XY \), \( XXZ \), and topological models such as the Su-Schrieffer-Heeger and Chern models \cite{Saubhik_Free_fermionic}. For the models explicitly considered, we found that for each system size, there exists a range of uncertainties in control parameters for which critical enhancement of metrological performance, as captured by QFI scaling, is preserved. Importantly, we showed that even though such uncertainties ultimately bound the maximal achievable precision, useful quantum advantage remains accessible at the critical point. This establishes that critical systems, while sensitive to nuisance parameters, are nonetheless robust enough to serve as viable probes for quantum-enhanced sensing.

In summary, our results provide a resolution to open questions  regarding the role of nuisance parameters in quantum metrology and constitute an important step toward understanding the practical limitations and fundamental noise sources in critical sensing protocols.


\acknowledgements
The authors are grateful to Victor Montenegro, Chiranjib Mukhopadhyay, and Andrew Mitchell for fruitful discussions and thank the organizers of QUMINOS. G.M. acknowledges support from Equal1 Laboratories Ireland Limited. SC acknowledges support from the John Templeton Foundation Grant ID 62422 and the Alexander von Humboldt Foundation. K.G. was supported by the Lise-Meitner Fellowship M3304-N of the Austrian Science Fund (FWF). 

\appendix

\section{Quantum Fisher information matrix for Landau-Zener toy-model}\label{A:LZ}

\begin{figure*}[htb!]
 	\centering
	\includegraphics[width=0.9\linewidth]{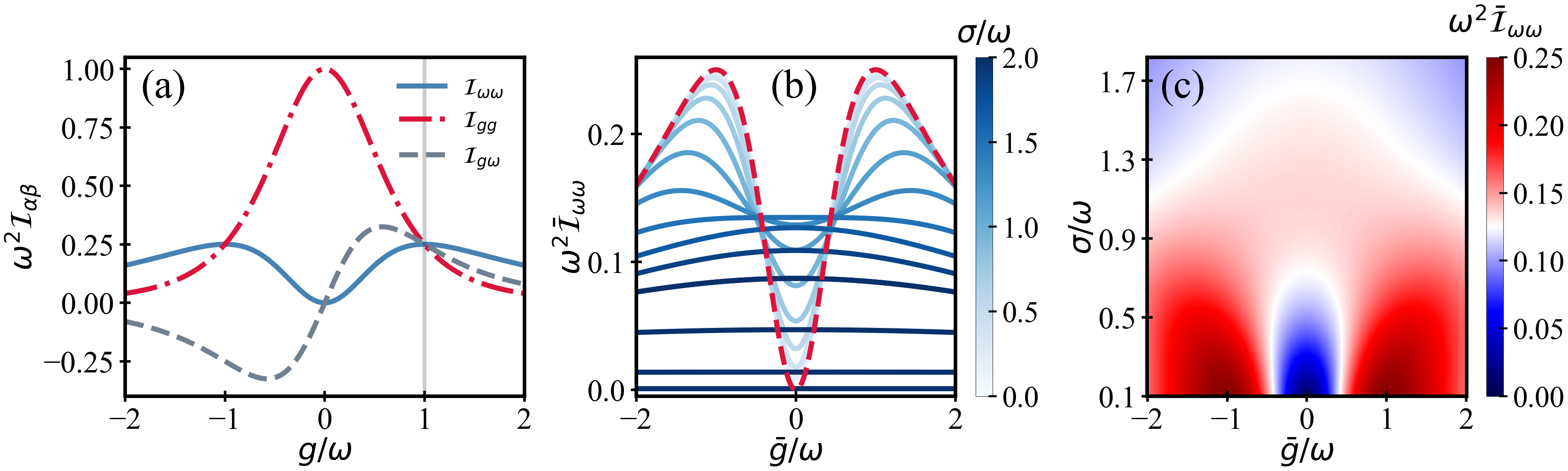}
\caption{\textit{Landau-Zener model:} (a) Elements of the QFIM for various $g$ when the bare frequency $\omega = 1$. (b) Sensitivity to the bare frequency $\omega$ with various degrees of uncertainty in the control parameter $g$. The dashed red line shows the ideal sensitivity ${\mathcal{I}}_{\omega\omega}^{SP}$ of the single parameter scenario when $g$ is known. The solid blue lines show attainable QFI ${\bar{\mathcal{I}}}_{\omega\omega}$ with uncertainty in the coupling $g$. (c) The full phase diagram in attainable QFI with varying degree of precision quantified by $\sigma$.  } 
	\label{fig:LZ_Elements_Uncertain}
\end{figure*}

The Landau-Zener model~\cite{zener1932non} describes a two-level system in a control field $g$,
\begin{align}\label{eq:LZ}
    \hat H_{\mathrm{LZ}} = \frac{\omega}2 \hat \sigma_z - \frac{g}2 \hat \sigma_x .
\end{align}
where $\omega$ is the level splitting for $g = \omega$, and $\hat \sigma_i$ is the $i$th Pauli matrix. Being a single-particle system, the Landau-Zener model does not exhibit a QPT. However, it features an avoided crossing and can therefore be considered as a toy model for criticality~\cite{zhang2009directcriticality,innocenti2020ultrafast,Gietka2021adiabaticcritical}. The instantaneous ground state of the Landau-Zener model is given by,
\begin{align}
    \label{eq:GS_LZ}
  |\mathrm{\psi_0}\rangle = - \frac{g+\sqrt{\omega^2+g^2}}{\sqrt{ 2g \left(g+\sqrt{\omega^2+g^2}\right)+  2\omega^2}}|\!\downarrow\,\rangle  \\ +\frac{\omega}{\sqrt{ 2g \left(g+\sqrt{\omega^2+g^2}\right)+ 2\omega^2}} |\!\uparrow\,\rangle.
\end{align}
Using Eq.~(3) from the main text, we can readily obtain the QFI matrix for the above ground state,
\begin{align}
{\hat{\mathcal{I}}}_{LZ} = 
    \begin{pmatrix}
\frac{g^2}{\left(g^2 + \omega^2\right)^2} &  -\frac{g \omega}{\left(g^2 + \omega^2\right)^2} \\
-\frac{\omega g}{\left(g^2 + \omega^2\right)^2} & \frac{\omega^2}{\left(g^2 + \omega^2\right)^2}
\end{pmatrix}.
\end{align}
The above QFI matrix is always singular regardless of the distance to the avoided crossing, i.e., $\rm{\det}\left[ \hat{\mathcal{I}}_{LZ} \right] = {\mathcal{I}}_{\omega\omega}{\mathcal{I}}_{g g} - {\mathcal{I}}_{\omega g}^2 = 0$. In Fig.~\ref{fig:LZ_Elements_Uncertain}(a) we show the elements of the QFIM as a function of driving parameter $g$ when $\omega = 1$, with ${\mathcal{I}}_{\omega\omega}$, ${\mathcal{I}}_{g g}$, and ${\mathcal{I}}_{\omega g}$ depicted in the solid blue, dash-dotted red, and dashed grey lines respectively. We note that the quantity $\omega^2 {\mathcal{I}}_{\alpha\beta}$ is a universal scaling function of the rescaled parameter $g/\omega$. In general, we may expect accidental crossings of the elements of the QFIM, in (a) we highlight the coalescence when $\omega = g = 1$. We attribute the singularity of the QFIM in the Landau-Zener model to the fact that we are not sensing separately parameters $\omega$ and $g$, but rather, an effective ratio of $g/\omega$. As we are sensing an effective ratio of parameters, precisely when $\omega = g$, the ratio $g/\omega = 1$, and elements of the QFIM attain the same value, and coalesce.

The multiparameter case represents the scenario when neither the bare frequency $\omega$ or the driving field $g$ are known. As in the main text, we investigate the scenario in which we wish to estimate the bare frequency $\omega$, but we have some finite window of resolution in the driving parameter $g$. This finite window of resolution represents the degree of uncertainty we have, quantified by the variance of a Gaussian distribution $\sigma$. The corresponding sensitivity is found viz.,

\begin{align}
    \label{Eq:Average_LZ}
    \hat{{\varrho}}\left(\omega,\bar g,\sigma\right) =  \int_{-\infty}^{\infty} d g \frac{\exp{-\frac{1}{2}\left(\frac{g - \bar{g}}{\sigma}\right)^2}}{\sqrt{2 \pi \sigma^2}} \hat{\varrho}\left(\omega, g\right),
\end{align}
with $\hat{\varrho}\left(\omega, g\right)$ the ground state density matrix of Eq.~\eqref{eq:GS_LZ}, $\sigma$ is the variance, and $\bar{g}$ is the expected value of the driving parameter $g$. In Fig.~\ref{fig:LZ_Elements_Uncertain} (b) the ideal sensitivity when the control field $g$ is perfectly known---i.e., the single parameter sensitivity, ${\mathcal{I}}_{\omega\omega}^{SP}$, is shown in the dashed red line. The solid blue lines depict the attainable sensitivity ${\bar{\mathcal{I}}}_{\omega\omega}$ for various uncertainty quantified by $\sigma$. We note that in the limit $\sigma \rightarrow \infty$, when the driving parameter $g$ is essentially unknown, there is zero attainable sensitivity to the bare frequency $\omega$. This is because the $\sigma \rightarrow \infty$ limit is essentially the multiparameter scenario: as the QFIM is singular, when the driving parameter $g$ is unknown, the frequency $\omega$ cannot be inferred. In scenarios where the QFIM is non-singular, we would expect the $\sigma \rightarrow \infty$ limit to yield some finite sensitivity. In panel (c) we show the full phase diagram for the attainable sensitivity when there is uncertainty in the driving parameter $g$.

\vspace{1cm}
\section{Quantum Fisher information matrix for transverse field Ising model}\label{A:TFIM}

In this section, we provide a simple prescription for finding a closed form solution for the QFIM and determinant of QFIM for any system that can be mapped to free fermions. We present these results by considering the TFIM as a paradigmatic example. The system can be diagonalized by first using the Jordan-Wigner transformation and then going to momentum space to obtain,
\begin{align}
\begin{split}
    \hat{H}_k = 2\sum_k\left(~\left[\omega - g\cos{k}\right]\left[\hat{c}_{k}^\dagger\hat{c}_{k}^{\phantom{\dagger}}  - \hat{c}_{-k}^{\dagger}\hat{c}_{-k}^{\phantom{\dagger}}\right]\right. \\ \left.-g\sin{k}\left[i e^{-2 i \phi} \hat{c}_{k}^{\dagger} \hat{c}_{-k}^{\dagger} - i e^{2 i \phi}\hat{c}_{k}^{\phantom{\dagger}}\hat{c}_{-k}^{\phantom{\dagger}}\right] ~\right),
\end{split}
\end{align}
where the arbitrary overall phase, $\phi$, was introduced in writing the model in the quasi-momentum $k$-space basis. The phase factor, $\phi$, simply corresponds to changing the phase of pairs created in momentum space. All Hamiltonian terms now come in momentum space pairs, $\left(k, -k\right)$, where one can define the full Hamiltonian by considering only the positive momentum $k$ values, corresponding to $k =  \pi\left(2 n +1\right)/N$ where $n = 0,1,2,..., N/2 - 1$. As noted in \cite{10.21468/SciPostPhysLectNotes.82}, the Hamiltonian's $\hat{H}_k$ live in a 4-dimensional space spanned by the states $\{\ket{0},\hat{c}_{k}^{\dagger}\ket{0},\hat{c}_{-k}^{\dagger}\ket{0},\hat{c}_{k}^{\dagger}\hat{c}_{-k}^{\dagger}\ket{0}\}$, which has a single non-trivial $2 \times 2$ block. To deal with the necessary combination of states living in these non-trivial blocks, spanned by $\{\ket{0},\{\hat{c}_{k}^{\dagger}\hat{c}_{-k}^{\dagger}\ket{0}\}$, we introduce fermionic two-component spinors
\begin{equation}
    \hat{\Psi}_k^{\phantom{\dagger}} = \begin{pmatrix}
        \hat{c}_{k}^{\phantom{\dagger}} \\ \hat{c}_{-k}^{\dagger}
    \end{pmatrix}, \qquad \hat{\Psi}_{k}^{\dagger} = \begin{pmatrix}
        \hat{c}_{k}^{\dagger} & \hat{c}_{-k}^{\phantom{\dagger}}
    \end{pmatrix},
\end{equation}
allowing us then to write the Hamiltonian of each momentum space block, $\hat{H}_k$, in the form
\begin{equation}
    \hat{H}_{k} = 2\Psi_{k}^{\dagger}\begin{pmatrix}
        \left(\omega - g \cos{k} \right) & -i e^{-2 i \phi} g \sin{k} \\ i e^{2 i \phi} g \sin{k} & -\left(\omega - g \cos{k} \right)
    \end{pmatrix}\Psi_{k}^{\phantom{\dagger}},
\end{equation}
where $\hat{H} = \sum_{k > 0}\hat{H}_k$. Making the common choice of phase, $\phi = - \pi/4$, corresponds to working in the $x$-$z$ pseudo-spin basis, allows us to write the Hamiltonian in a more familiar manner viz. the pseudo-spin Pauli matrices as,
\begin{equation}
    \hat{H}_{k}^{spin} = 2\Psi_{k}^{\dagger}\left[ \left(\omega - g \cos{k}\right)\sigma_z + \left(g\sin{k}\right)\sigma_x \right]\Psi_{k}^{\phantom{\dagger}}.
\end{equation}
Solving the $2 \times 2$ eigenvalue problem for the pseudo-spin Hamiltonian provides us with the ground state energy of each momentum-space block,
\begin{equation}
    \epsilon_k = -2\sqrt{g^2 + \omega^2 - 2 g \omega \cos{k}},
\end{equation}
with a corresponding (un-normalised) ground state solution of each block given by
\begin{equation}
    \label{Eq:Appendix_GS_k}
    \ket{\psi_{GS}}_k = \cos{\frac{\theta_k}{2}}\ket{0}_k + \sin{\frac{\theta_k}{2}}\ket{1}_k,
\end{equation}
where the angle is defined as $\tan{\theta_k} = \frac{g \sin{k}}{g \cos{k} -  \omega}$. The ground state for an $N$-site system is simply given as the tensor product $\ket{\psi_{GS}}_N = \bigotimes_{k>0} \ket{\psi_{GS}}_k$ of the ground state of each momentum space block $\hat{H}_k$. The QFI is additive under tensor product and for a pure state of the form $\ket{\Psi} = \ket{\psi}_A \otimes \ket{\psi}_B \otimes ...., \ket{\psi}_N$ can be decomposed as,
\begin{equation}
    \label{eq:Pure_State_Tensor}
    {\mathcal{I}}_{\alpha\beta}\left[|\Psi\rangle\right] = {\mathcal{I}}_{\alpha\beta}\left[ \ket{\psi}_A \right] + {\mathcal{I}}_{\alpha\beta}\left[ \ket{\psi}_B \right] + ..., {\mathcal{I}}_{\alpha\beta}\left[ \ket{\psi}_N \right]
\end{equation}
with ${\mathcal{I}}_{\alpha\beta}\left[ \ket{\psi}_i \right]$ defined in Eq.~(3). By exploiting the additive property of the QFI under tensor product and the fact that the ground state of a free fermion system can be written as $\ket{\psi_{GS}}_N = \bigotimes_{k>0} \ket{\psi_{GS}}_k$, elements of the QFIM can readily be obtained 
\begin{equation}
    {\mathcal{I}}_{\alpha\beta} = 4 \sum_k \left[\langle \partial_\alpha \psi_{GS}|\partial_\beta \psi_{GS}\rangle_k -\langle \partial_\alpha \psi_{GS}|\psi_{GS}\rangle_k \langle \psi_{GS}| \partial_\beta \psi_{GS} \rangle _k\right],
\end{equation}
as the sum of the QFI of each momentum space block. Alternatively, the angle $\theta_k$ containing information regarding the unitary transformation required to diagonalize the system can be used to compute elements of the QFIM viz.,
\begin{equation}
    {\mathcal{I}}_{\alpha\beta} = \sum_k \partial_\alpha \arctan{\theta_k} \times \partial_\beta \arctan{\theta_k}.
\end{equation}
Equipped with this, the QFIM can readily be obtained for arbitrary system size $N$ of the TFIM as,
\begin{align}
  \hat{\mathcal{I}} =\sum_{k} \begin{pmatrix}
    \frac{g^2 \sin^2{k}}{\left(g^2 + \omega^2 - 2 g \omega \cos{k} \right)^2} & -\frac{g \omega \sin^2{k}}{\left(g^2 + \omega^2 - 2 g \omega \cos{k} \right)^2} \\ -\frac{g \omega\sin^2{k}}{\left(g^2 + \omega^2 - 2 g \omega\cos{k} \right)^2} & \frac{\omega^2 \sin^2{k}}{\left(g^2 +\omega^2 - 2 g \omega \cos{k} \right)^2}
\end{pmatrix}, 
\end{align}
with $k =  \pi(2n+1)/N$ and $n = 0,1,2,\ldots,\frac{N}2 - 1$, where at the critical point of the TFIM, $g=\omega$, elements of the above QFI matrix follow a triangular sequence and coalesce to $(N^2+N)/8\omega^2$. The determinant of the QFIM follows as determinant of the sum of the QFIM of each momentum space block,
\begin{equation}
    \rm{det}\left[ \hat{\mathcal{I}}\right] = \rm{\det}\left[\sum_{k} \begin{pmatrix}
    \frac{g^2 \sin^2{k}}{\left(g^2 + \omega^2 - 2 g \omega \cos{k} \right)^2} & -\frac{g \omega \sin^2{k}}{\left(g^2 + \omega^2 - 2 g \omega \cos{k} \right)^2} \\ -\frac{g \omega\sin^2{k}}{\left(g^2 + \omega^2 - 2 g \omega\cos{k} \right)^2} & \frac{\omega^2 \sin^2{k}}{\left(g^2 +\omega^2 - 2 g \omega \cos{k} \right)^2}
\end{pmatrix} \right],
\end{equation}
which in the case of the QFIM is singular for all $N$, but in full generality is not strictly singular for any free fermion system and set of parameters. 

\newpage

\section{Additional figures}\label{A:C}
\begin{figure}[htb!]
 	\centering
	\includegraphics[width=1\linewidth]{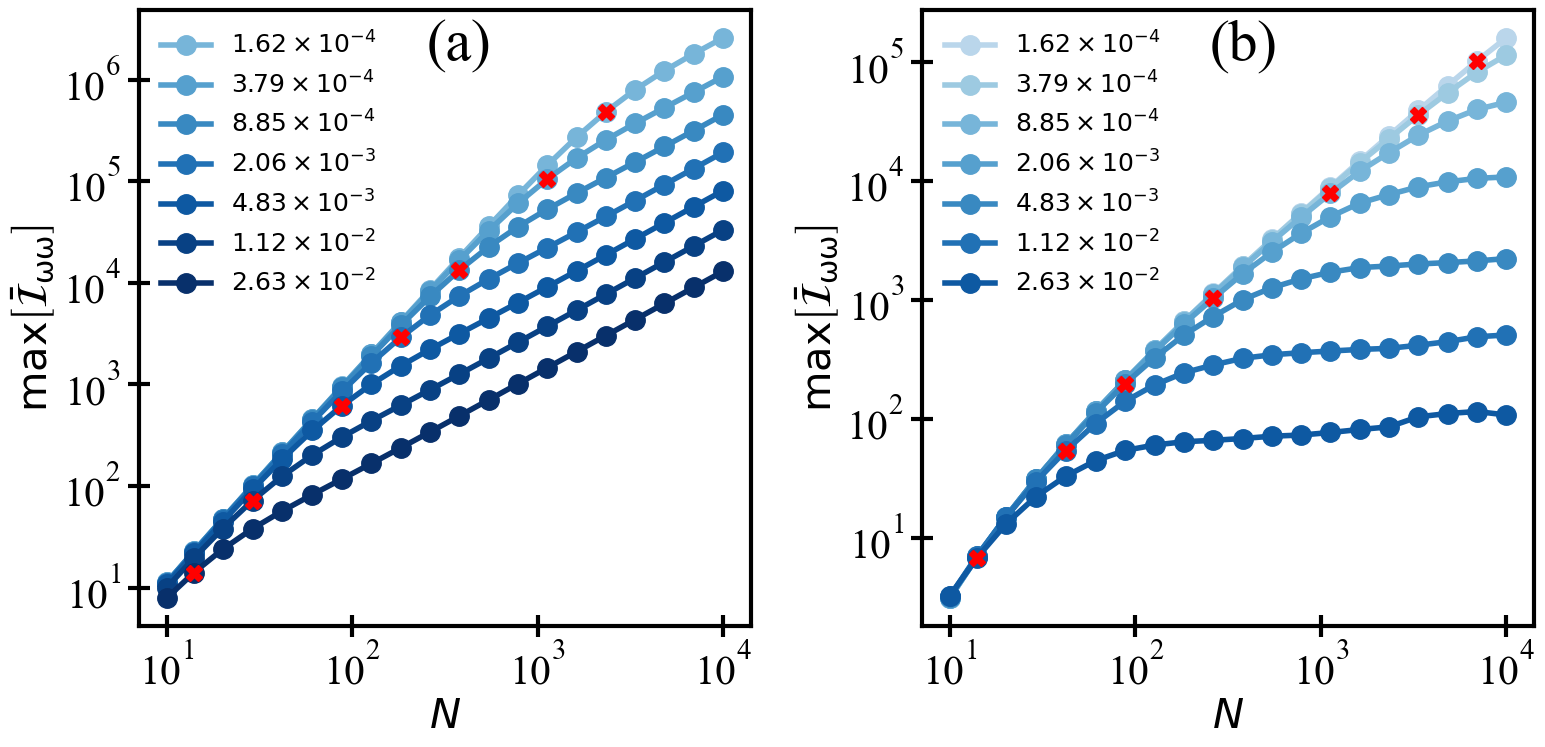}
\caption{Scaling of the peak QFI ${\bar{\mathcal{I}}}_{\omega\omega}$, with a given uncertainty $\sigma/\omega$ in the control parameter $g$ in shades of blue, as a function of system size $N$ in the TFIM (a) and LMG (b). Red crosses denote the value at which the QFI \emph{feels} the uncertainty of the control parameter $g$ and shows the full compliment of data within the relevant uncertainty range used to generate panels (d) and (h) of Fig. 1 from the main text.}
	\label{fig:Uncertain_data}
\end{figure}

For the sake of clarity, in Figs~1(b) and (f) from the main text, we presented only part of the data used to generate Figs~1(d) and (h). Here we provide all the relevant generated data and plot them in Fig.~\ref{fig:Uncertain_data}. Note that because of the finite resolution in $\sigma$, the red crosses in Figs~1(d) and (h) appear bumpy.

%

\end{document}